\documentstyle[12pt,psfig]{article}
\def\be{\begin{equation}}
\def\ee{\end{equation}}

\def\TL{\hfil$\displaystyle{##}$}
\def\TR{$\displaystyle{{}##}$\hfil}

\def\TT{\hbox{##}}
\def\seqalign#1#2{\vcenter{\openup1\jot
  \halign{\strut #1\cr #2 \cr}}}

\def\fixit#1{}

\def\mop#1{\mathop{\rm #1}\nolimits}
\def\tr{\mop{tr}}
\def\Tr{\mop{Tr}}

\def\href#1#2{#2}  

\def\eqalign#1{\vcenter{\openup1\jot
    \halign{\strut\span\TL & \span\TR\cr #1 \cr
   }}}
\def\lbldef#1#2{\expandafter\gdef\csname #1\endcsname {#2}}
\def\eqn#1#2{\lbldef{#1}{(\ref{#1})}%
\begin{equation} \eqalign{#2} \label{#1} \end{equation}}

\begin{document}
\baselineskip=16pt
\pagestyle{plain}
\setcounter{page}{1}

\begin{titlepage}

\begin{flushright}
hep-th/0204051 \\
PUPT-2029
\end{flushright}
\vfil

\begin{center}
{\huge A semi-classical limit \\[10pt]
of the gauge/string correspondence}
\end{center}

\vfil
\begin{center}
{\large S. S. Gubser,\footnote{E-mail: 
{\tt ssgubser@princeton.edu}}
I. R. Klebanov\footnote{E-mail: \tt klebanov@feynman.princeton.edu}
and
A. M. Polyakov\footnote{E-mail: \tt polyakov@princeton.edu}
}\end{center}

$$\seqalign{\span\TL\, & \sl\span\TT}{
 & Joseph Henry Laboratories, Princeton University, 
 Princeton, NJ  08544, USA  \cr
}$$
\vfil

\begin{center}
{\large Abstract}
\end{center}

 A world-sheet sigma model approach is applied to string theories dual
to four-dimensional gauge theories, and semi-classical soliton
solutions representing highly excited string states are identified
which correspond to gauge theory operators with relatively small
anomalous dimensions.  The simplest class of such states are strings
on the leading Regge trajectory, with large spin in $AdS_5$.  These
correspond to operators with many covariant derivatives, whose
anomalous dimension grows logarithmically with the space-time spin.
In the gauge theory, the logarithmic scaling violations are similar to
those found in perturbation theory.  Other examples of highly excited
string states are also considered.

\vfil
\begin{flushleft}
April 2002
\end{flushleft}
\end{titlepage}
\newpage
\renewcommand{\thefootnote}{\arabic{footnote}}
\setcounter{footnote}{0}
\renewcommand{\baselinestretch}{1.2}  
\tableofcontents

\section{Introduction}
\label{Introduction}

It is by now well established that gauge theories can be described by
strings, representing their flux lines. These strings, however,
are unusual. In the case of the pure glue gauge theory they propagate
in a five dimensional warped space 
\cite{Sasha} and, if some other fields
are present, even in higher dimensions. The string Lagrangian is
that of the non-linear sigma model with the target space
coinciding with the one described above. The simplest and best
explored case of this gauge/string duality is presented by the
Yang-Mills theory with 
${\cal N}=4$ supersymmetry. Conformal invariance
and other symmetries of this theory unambiguously determine the
geometry of the warped target space. It was found in 
\cite{juanAdS,gkPol,witHolOne},
developing some earlier findings of \cite{absorb}, 
that the desired string
theory in this case lives in the space $AdS_5 \times S^5$ and
that there is a unique prescription relating physical quantities
in the string and gauge pictures. Many more complicated examples
have been analyzed since then, confirming the existence
of a dual string picture for various gauge theories.

However, until very recently, it is mostly the supergravity limit
(corresponding to the one-loop approximation of the sigma model) that 
has been tested, basically because of the challenges inherent in 
formulating the appropriate sigma models\footnote{See \cite{Berkovits} and
references therein for the most highly developed approach to this problem.}
and of solving them once 
they are formulated.
It has been recently pointed out in \cite{polWord} 
that for gauge theory
operators with very high bare dimension, such as operators with high
spin, R-charge, etc., the supergravity approximation is inadequate even
at small curvatures of the AdS (corresponding to large
't~Hooft coupling $\lambda = g_{\rm YM}^2 N$). 
It is expected that for for certain classes of such operators
the dimensions asymptotically become integer spaced,
even when $\lambda \gg 1$.

These expectations were confirmed and considerably strengthened in 
\cite{bmn} for a
special class of operators with high R-charge. 
It turned out to be possible
in this case to approximate the AdS space by the much simpler and exactly
solvable pp-wave \cite{Metsaev} by taking the Penrose limit 
\cite{papa}. What remained unclear was
whether this success is 
a manifestation of a general mechanism.

In this paper we will argue that the gauge theory states with large
quantum numbers are described by certain solitons of the nonlinear sigma
model. The reason for this is easy to explain. The world sheet dimensions of
the 2d conformal field theories are expressed in the radial quantization
scheme as eigenvalues of the hamiltonian $H=L_{0}+\overline{L_{0}}.$ For the
small sigma model coupling, the spectrum of such a hamiltonian contains the
usual perturbative oscillations which are effectively described in the
target space by the supergravity approximation.

However in some cases (which we discuss below) there are also soliton
states with world sheet dimensions $\sim \frac{1}{\alpha }$ where
$\alpha =1/\sqrt\lambda$ 
is the sigma model coupling. Discussion of these ``conformal
solitons'' and their gauge counterparts is the main content of the
present paper.  There is already a well known application of
classical solutions: the calculation of the Wilson loop
\cite{juanWilson}.  For closed loops, the relevant solutions
(minimal surfaces in $AdS_5$) are instanton-like, having a finite
action once a divergence associated to point-like charges is
regulated.  Our solutions, by contrast, are solitons, and they have
different physical meaning.  They describe operators with high bare
dimensions (very long ``words'' in the terminology of \cite{polWord}).
Similar operators describe asymptotic behavior of the deep inelastic
scattering.

An important result, which is responsible for some of the
 phenomenological successes of perturbative QCD in deep inelastic
scattering, 
 is the logarithmic dependence of
anomalous dimension on spin for high-spin gauge theory operators
\cite{GrossWilczek}.  With the (relatively) 
new techniques of AdS/CFT correspondence
in hand \cite{juanAdS,gkPol,witHolOne} (for a review see
\cite{MAGOO}), we might ask how (and whether) 
logarithmic scaling violations might
arise in a regime where stringy techniques in anti-de Sitter space are
well-controlled.  On the field theory side, such effects need not be
tied to a non-zero beta-function: anomalous dimensions of non-chiral
operators can arise even in a superconformal theory like ${\cal N}=4$
super-Yang-Mills.  On the string theory side, it might seem difficult
to make progress, because gauge theory operators which are neither
chiral nor descendants of chiral operators are thought to correspond
to excited string states rather than supergravity modes.  Such states
have masses at least of order $1/\sqrt{\alpha'}$, so their dimensions
typically grow as $\lambda^{1/4}$ \cite{gkPol,witHolOne}. 
Full control of the string spectrum in $AdS_5
\times S^5$ and the operator-state map seems to be beyond our grasp at
present.  However, we will argue that certain features of the string
spectrum can be well-understood from a 
non-linear sigma-model perspective,
which bypasses (or at least postpones) the usual difficulty with
Ramond-Ramond backgrounds.

We will consider a class of gauge invariant
operators with high Lorentz
spin $S$. These operators are quite different from those with
large R-charge: their presence does not rely on the supersymmetry
of the gauge theory. In fact, they are well-known to be present
in QCD where they were studied early on in the context of deep
inelastic scattering 
\cite{GrossWilczek}. Analogous operators are present in the
${\cal N}=4$ supersymmetric $SU(N)$ gauge theory, for example,
\eqn{TwistTwo}{
   \tr X^I \nabla_{(\mu_1} \cdots 
     \nabla_{\mu_s)} X^I \,.
  }
 In free field theory, such an operator would have ``twist two:'' that
is, its dimension is $S+2$, while its space-time spin is $S$.  Similar
operators play an important role in deep inelastic scattering, where
leading weak coupling
corrections of the form $\Delta = (S+2) + c_1 g_{YM}^2 N \ln S$ can
be probed experimentally.
 Unlike the
operators with large R-charge, such high-spin operators cannot
be chiral or `nearly chiral,' hence their anomalous dimensions
would seem to be difficult to control at strong coupling.
Nevertheless, for sufficiently large $S$, we present a simple
semi-classical method for calculating them. 
Our approach involves finding a soliton
solution of the $\sigma$-model describing type IIB strings on
$AdS_5\times S^5$. Although a solution of such a RR 
$\sigma$-model is lacking, finding the solitons describing
the high-spin closed strings is a much easier problem.

Consider, for example, closed strings on the leading Regge trajectory,
i.e.{} those with the smallest mass for a given spin $S$.
It is well known that an approximate description of such states
in flat space is a folded closed string which spins as a rigid
rod around its center. To study the gauge theory we simply
replace the flat space by the global $AdS_5$ metric,
\begin{equation}
\label{global}
ds^2 = R^2 \left (
-dt^2 \cosh^2\rho +d\rho^2 +
\sinh^2 \rho d\Omega_3^2 \right )
\ ,
\end{equation}
 and consider a spinning folded closed string whose center lies at
rest at $\rho=0$.\footnote{We are grateful to Juan
Maldacena for suggesting this to us. In the AdS$_4$ case such
a solution was constructed and studied in \cite{Vega}.}
We take the usual large $N$ `t Hooft limit 
keeping $\lambda = g_{\rm YM}^2 N$ fixed, and make $\lambda$
very large so that the curvature of the metric is small.
For spin $S \ll \sqrt \lambda$ we show that the standard 
AdS/CFT analysis of operator dimensions 
\cite{gkPol,witHolOne} applies. For $S\sim \sqrt\lambda$, however,
this analysis becomes inadequate because the string gets
appreciably stretched across AdS$_5$.
Identifying the energy in the global coordinates
with the conformal dimension in the dual gauge theory, we find for
$S\gg \sqrt\lambda$,
  \eqn{StateDeltaS}{
   \Delta - S = {\sqrt\lambda\over \pi} \ln (S/\sqrt\lambda)
   + O(S^0) \,.
  }
 The result \StateDeltaS\ is interesting because 
these anomalous dimensions
grow logarithmically with spin only in gauge theories \cite{GrossWilczek}.
 Details of the calculation leading to \StateDeltaS\ will be provided
in section~\ref{Regge} (the logarithmic growth
of $\Delta - S$ on the AdS side is closely related 
to the $\ln S$ scaling of the 
proper length of the string for large $S$).  
The result \StateDeltaS\ shows that for these states the
anomalous dimension is small compared to the bare one:
  \eqn{StillSmall}{
   {\Delta - S\over S} \sim { \ln (S/\sqrt\lambda)\over S/\sqrt\lambda }
    \ll 1 \,.
  }
 Comparison with the perturbative results motivates 
the following formula for the leading term in the dimension of
a high-spin operator on the leading Regge trajectory:
\be \label{gist}
\Delta - S =f (\lambda) \ln S
\ ,
\ee
where $f(\lambda)= a_1 \lambda + a_2\lambda^2 + \ldots$ 
is a certain function of the 't Hooft coupling.
Indeed, in many gauge theories (including non-supersymmetric ones!)
it has been argued that the leading term in the anomalous dimensions
grows as $\ln S$ to all orders in perturbation theory, and probably
also non-perturbatively \cite{Korch}.\footnote{
We are grateful to G. Korchemsky for bringing these papers
to our attention, and for helpful comments after the original
version of this paper appeared.} The analysis of \cite{Korch}
is based on studying Wilson loops with cusps, and
the function $f(\lambda)$ appears in this approach
as the ``cusp anomalous dimension.'' 
Therefore, the formula (\ref{gist}) may be a new point of contact
between gauge theory and string theory. 

Although we discuss operators with high Lorentz spin in the most detail,
we believe that our approach of identifying highly excited gauge theory
operators with $\sigma$-model solitons is quite general and can be applied 
to other kinds of operators. Our techniques may also be generalized to
backgrounds describing non-conformal gauge theories.

The organization of the paper is as follows.  In
section~\ref{RCharge}, we re-examine the case of large R-charge,
obtaining the results of \cite{bmn} from a different perspective,
developing ideas in \cite{polWord}.  In section~\ref{Regge}, we use
similar methods to treat spinning strings in $AdS_5$.  In
section~\ref{Others} we study some other examples of highly excited
strings.  And in section~\ref{Perturbative} we review the appearance
of logarithmic scaling violations in perturbative gauge theory and
discuss the relation with the results of section~\ref{Regge}.

\section{Large R-charge revisited}
\label{RCharge}

Our work is largely inspired by the recent important
observation that, if the bare dimension of an operator
becomes large simultaneously with $\lambda$, then the standard
rule that the anomalous dimension blows up as $\lambda^{1/4}$
may be violated \cite{polWord,bmn}. A simple example of such
a modification is provided by certain operators whose
R-charge $J$ is of order $\sqrt\lambda$. In \cite{bmn}
it was shown that such operators are in one-to-one correspondence
with {\it all} type IIB string states in a RR-charged pp wave
background which is a Penrose limit of $AdS_5\times S^5$
\cite{papa}.
Exact solvability of string theory in this pp wave background
\cite{Metsaev},
together with the AdS/CFT correspondence, predict that the dimensions
of these operators are given by
\be \label{ppdim}
\Delta = J +\sum_{n=-\infty}^\infty N_n \sqrt {1 + {\lambda n^2\over
J^2} }
\ ,
\ee
where $N_n$ is the excitation number of the $n$-th string mode.
The operators with $N_n=0$ for all $n\neq 0$ are chiral while
all the rest are non-chiral. Nevertheless, in the limit
$\lambda\rightarrow\infty$, where $J^2/\lambda$ is held fixed,
it is found that the anomalous dimension becomes
negligible compared to the bare dimension,
$(\Delta - J)/J \rightarrow 0$.

In this section we will reconsider the case of the large R charge.
We will treat it differently (but with the same results) in
order to set the stage for the more complicated cases. The bosonic part of
the action describing the AdS$_{5}\times S_{5}$ sigma model is given by
  \begin{equation} \label{saction}
   S=\frac{1}{2\alpha }\int d^2\sigma \, \sqrt{g} ((\nabla_\alpha n)^{2}+
   (\nabla_\alpha N)^{2})+...
  \end{equation}
where $n$ is a unit vector describing $S^5$ and $N$ is a hyperbolic unit
vector describing $AdS_5$.  The dots stand for the fermionic and RR terms.
The sigma model coupling $\alpha $ is related to  
$ \lambda $ as $\alpha =\frac{1}{\sqrt{\lambda }}.$ 
Let us consider the S$ _{5}$ part first. 
The R charge $J$ is the angular momentum on $S^5$.  The
standard perturbative formula for the world sheet dimension $\delta $ gives
for large $J$
\begin{equation} \label{dim}
\delta =\frac{1}{2}\alpha J^{2}
\end{equation}
As was explained in \cite{polWord}, 
the AdS part gives the same contribution but with
the reversed sign (because of the negative curvature) and thus
\begin{equation}
\delta =\frac{1}{2}\alpha (J^{2}-\Delta ^{2})
\end{equation}
where $\Delta $ is the space-time dimension.
The mass shell condition is given by $\delta =-l+1,$ 
where $l$
is the ``level,'' that is the number of the world sheet derivatives in the
corresponding operator. From this we see that there should exist gauge
theory operators with 
$\Delta$ and $J$ of order $1/\alpha$ but with 
$\Delta -J$ of order $1$.
However, we have to correct the above formula by 
accounting for the terms in
the world-sheet dimension containing the powers of 
$\alpha J\sim 1,$ while
neglecting the powers of $\alpha .$ 
In order to fix the angular momentum we
will consider the hamiltonian in the rotating frame $\widetilde{H}=H-\omega $
$J$ and look for the minimal energy. This procedure is equivalent to
considering the classical solution for the $S^5$ sigma model
 \begin{equation} \label{triv}
\theta =0\,,\qquad \psi =\omega \tau \,,
\end{equation}
 where $\theta $ and $\psi $ are the polar and azimuthal angles on
$S^2$ (we disregard other angles of $S^5$ , which later will
give some trivial modification of our formulae); we parametrize
the world sheet by the variables $\tau $ and $\sigma$.
Geometrically this solution represents a closed string 
collapsed to a point which rotates around the equator. The 
world sheet energy
of the above solution is given by
$\frac{1}{2\alpha }\omega ^{2}$
which gives the formula (\ref{dim}) after using the relation 
$J=\frac{\partial E}{%
\partial \omega }$. To account for the corrections
mentioned above, it is sufficient
to consider harmonic oscillations near this solution. The Lagrangian for the
$\theta $ oscillations, as seen from (\ref{saction}), has the form
 \begin{equation}
\alpha L=\frac{1}{2}
\left [ (\nabla \theta )^{2}+\omega ^{2}\cos ^{2}\theta \right ]\simeq
\frac{1}{2} \left [
\left( \nabla \theta \right) ^{2}-\omega ^{2}\theta ^{2}+\omega^2 \right ]
\,.\end{equation}
 {}From here we derive the oscillator spectrum of anomalous dimensions
of the spherical sigma model
 \begin{equation}
\delta =\frac{\alpha}{2 }J^{2}+\sum_{n}N_{n}\sqrt{n^{2}+\alpha ^{2}J^{2}}
\end{equation}
where we used the relation  $J=\omega/\alpha$.  The $N_{n}$
here is the excitation number of the n-th oscillator. The formula
for AdS part is obtained by the change $\alpha $ to $-\alpha$.
The on-shell condition
\begin{equation}
\delta (S^5)+\delta (AdS_5)\approx 0
\end{equation}
gives the formula (\ref{ppdim}),
which was previously derived in \cite{bmn} 
by the use of the Penrose limit. We see
that from the sigma model point of view it corresponds to a harmonic
approximation to a relatively 
trivial classical solution (\ref{triv}). We shall now
proceed to the more complicated solutions, describing  operators with high
spin.

\section{Leading Regge Trajectory}
\label{Regge}

In this section we consider a spinning closed string in $AdS_5$ to
understand the relation between dimension and spin for leading Regge
trajectory closed strings. 
To carry out this calculation it is
convenient to use the global $AdS_5$ metric (\ref{global}), so that
the energy is identified with the conformal dimension in the dual CFT.
The string is at the equator of $S^3$ and the azimuthal angle depends
on time:
  \eqn{phit}{
   \phi = \omega t \,.
  }
In the Nambu action for the string we pick a gauge where $\tau = t$
and $\rho$ is a function of $\sigma$.  The Lagrangian becomes
 \begin{equation} L = - 4 {R^2\over 2\pi \alpha'}
\int_0^{\rho_0} d\rho \, \sqrt{\cosh^2\rho - (\dot \phi)^2\sinh^2 \rho }
\end{equation}
 The maximum radial coordinate is $\rho_0$, and the factor of 4 comes
since there are four segments of the string stretching from 0 to
$\rho_0$, which is determined by
 \begin{equation}\coth^2 \rho_0 = \omega^2\ .
\end{equation}

The energy and the spin of the string are
\begin{equation} E =  4 {R^2\over 2\pi \alpha'}
\int_0^{\rho_0} d\rho \,
{\cosh^2\rho \over \sqrt{\cosh^2\rho - \omega^2\sinh^2 \rho } }
\label{EnergyS}\end{equation}
\begin{equation} S =  4 {R^2\over 2\pi \alpha'}
\int_0^{\rho_0} d\rho \, 
{\omega \sinh^2\rho \over \sqrt{\cosh^2\rho - \omega^2\sinh^2 \rho }
}
\label{SpinS}\end{equation}

The same expressions can also be derived in the conformal gauge
where the world sheet action is
\begin{equation} S={1\over 4 \pi \alpha'}\int d\tau d\sigma \, G_{ij}
\partial_\alpha X^i \partial^\alpha X^j
\ .
\end{equation}
 Now we need to impose the conditions
  \eqn{PMContraints}{
   T_{++} &= \partial_+ X^i \partial_+ X^j G_{ij} = 0  \cr
   T_{--} &= \partial_- X^i \partial_- X^j G_{ij} = 0 \ .
  }
 Inserting $t =e \tau$, $\phi =e \omega \tau$, $\rho = \rho(\sigma)$
into these equations, we find
 \begin{equation}
 \label{radial} (\rho')^2 = e^2 (\cosh^2 \rho - \omega^2 \sinh^2 \rho)\ .
\end{equation}
 Thus,
 \begin{equation} d\sigma = {d\rho \over e \sqrt{
\cosh^2 \rho - \omega^2 \sinh^2 \rho } }
\ .
\end{equation}
 We may now adjust $e$ so that the period of $\sigma$ is $2\pi$.

The space-time energy is given by
  \eqn{EnergyC}{
   E = {R^2\over 2 \pi\alpha'} e \int_0^{2\pi} d\sigma
    \, \cosh^2 \rho \,,
  }
 and the spin by
  \eqn{SpinC}{
   S = {R^2\over 2 \pi\alpha'} e\omega
    \int_0^{2\pi} d\sigma \, \sinh^2 \rho \,.
  }
 Changing the integration variable from $\sigma$ to $\rho$ we find the
previously derived expressions.

Since $R^4=\lambda \alpha'^2$, where $\lambda$ is the 't Hooft
coupling, these expressions specify $E/\sqrt \lambda$ and $S/\sqrt
\lambda$ as functions of $\omega$.  Therefore, the dependence of $
E/\sqrt\lambda$ on $S/\sqrt \lambda$ is known in parametric form.
Actually, the integrals in (\ref{EnergyS}) and (\ref{SpinS}) can be
expressed in terms of elliptic or hypergeometric functions, as we
shall explain in section~\ref{ArbitraryH}.  Here it will suffice to
give approximate expressions in the limits where the string is much
shorter or much longer than the radius of curvature of $AdS_5$.

{\bf Short strings}.  For large $\omega$, $\rho_0\approx 1/\omega$.
Here the string is not stretched much compared to the radius of
curvature of $AdS_5$, so we can approximate $AdS_5$ by flat metric
near the center. The calculation reduces to the standard spinning
string in flat space, and we get
 \begin{equation} E= {R^2\over \alpha' \omega}
\ ,\qquad 
S= {R^2\over 2\alpha' \omega^2}
\ ,
\end{equation}
so that 
\begin{equation}
\label{stand} E^2 = R^2 {2 S\over \alpha'}
\ .
\end{equation}
Using the AdS/CFT correspondence,
we have $\Delta = E$.
For large $\omega$, $S\ll \sqrt\lambda$.
In this regime we find agreement with the
AdS/CFT result
\begin{equation}\Delta^2 \approx m^2 R^2
\ .
\end{equation}
Indeed, for the leading closed string Regge trajectory, 
$m^2 = {2 (S-2)\over \alpha'}$, so the precise factor 
in (\ref{stand}) agrees with the AdS/CFT formula.

{\bf Long strings.}  The situation where $S\gg \sqrt \lambda$  corresponds
to $\omega$ approaching 1 from above:
 \begin{equation}\omega = 1+ 2 \eta \,,
\end{equation}
 where $\eta \ll 1$.  Then $\rho_0$ becomes very large, so that the string
feels the metric near the boundary of AdS:
 \begin{equation}\rho_0 \to {1\over 2} \ln (1/\eta)
\ .
\end{equation}
 Expanding the integrals for $E$ and $S$, one finds
 \begin{equation} E= { R^2\over 2\pi\alpha'} 
 ({1\over \eta} + \ln (1/\eta) +\ldots )
\ ,
\label{ELong}\end{equation}
\begin{equation} S= { R^2\over 2\pi\alpha'} 
({1\over \eta} - \ln (1/\eta) +\ldots )
\ .
\label{SLong}\end{equation}
 It follows that $E-S$ does
not approach a constant in the limit $S\gg \sqrt{\lambda}$, but instead
behaves as follows:
 \begin{equation} E- S = {\sqrt{\lambda}\over \pi} \ln (S/\sqrt \lambda)
+ \ldots
\label{logES}\end{equation}
Thus, the
situation is different from the large R-charge limit: the dependence
of the dimension on large spin is more complicated than for large
R-charge.  The logarithmic asymptotics in (\ref{logES}) 
are qualitatively the same as in perturbative
gauge theories, a topic which we will discuss in
section~\ref{Perturbative}.

It may be necessary to take the string coupling very small in order to
suppress decay of the long spinning strings into a number of shorter
strings corresponding to operators which are chiral primaries or
descendants of them. In the strict `t Hooft limit,
$g_{\rm str}\rightarrow 0$, hence these states are stable.

\begin{figure}
   \centerline{\psfig{figure=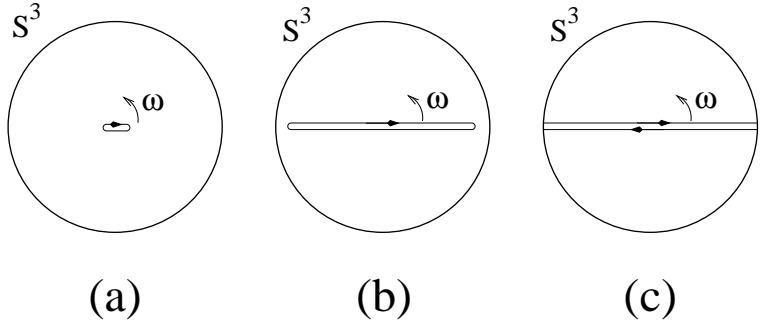,width=4in}}
   \caption{Short, long, and infinite strings spinning in
$AdS_5$, shown here in cross-section.}\label{figA}
  \end{figure}

{\bf Infinite strings:} Although energy and angular momentum diverge
as $\omega \to 1$ from above, one may nevertheless consider solutions
of the form \phit\ with $\omega < 1$: such a solution represents two
Wilson loops stretching in opposite directions, with the ends rotating
on the equator of $S^3$ with velocity $\omega$.  See
figure~\ref{figA}.

\section{Other macroscopic string states}
\label{Others}

The approach used in section~\ref{Regge} to treat strings spinning in
$AdS_5$ can also be used to approximately describe other highly
excited string states.  In sections~\ref{NorthPole}
and~\ref{Oscillate}, we present two examples.  In
section~\ref{ArbitraryH}, as a formal exercise, we extend the
discussion to states in the sigma model which do not necessarily
satisfy the on-shell conditions.

\subsection{String Spinning on $S^5$}
\label{NorthPole}

A similar calculation may be done for a string whose center of mass is
not moving on $S^5$, but which spins around that point and is
correspondingly stretched.  We lose nothing by assuming that this
point is the north pole.  Let us write the $S^5$ metric as
 \begin{equation} 
ds^2 = d\theta^2 + \sin^2\theta d\phi^2 +\cos^2 \theta d\Omega_3^2
\ ,
\end{equation}
 and adopt the ansatz $t =e\tau$, $\phi= e\omega \tau$,
$\theta = \theta(\sigma)$.
Then from $T_{++}= T_{--} = 0$ we get
  \eqn{thetap}{
   (\theta')^2 = e^2 (1 - \omega^2 \sin^2\theta) \,.
  }
Thus,
\begin{equation} d\sigma = {d\theta \over e \sqrt{
1 - \omega^2 \sin^2 \theta } }
\ .
\end{equation}
We may now adjust $e$ so that the period of $\sigma$ is $2\pi$.
The folded closed string is stretched along the $\theta$ direction
up to $\theta_0$ given by
\begin{equation}\sin \theta_0 = {1\over \omega}
\ .
\end{equation}

The space-time energy is given by
\begin{equation} E = {R^2\over 2 \pi\alpha'} e \int_0^{2\pi} d\sigma 
= 4 {R^2\over 2 \pi \alpha'}  
\int_0^{\theta_0} d\theta \, \left 
(1 - \omega^2 \sin^2\theta \right )^{-1/2}\ ,
\end{equation}
and the R-charge by
\begin{equation} J = {R^2\over 2 \pi\alpha'} e\omega
\int_0^{2\pi} d\sigma \, \sin^2 \theta =
4 {R^2\over 2 \pi \alpha'}  
\omega \int_0^{\theta_0} d\theta \, \left(
1 - \omega^2 \sin^2\theta \right )^{-1/2}
 \sin^2\theta
\ .
\end{equation} 
 These integrals can be expressed in terms of elliptic or
hypergeometric functions, but it would add little to the discussion to
present the explicit forms.  As before, the interesting regime is when
$\omega$ approaches $1$ from above, $\omega = 1 + 2\eta$. As $\eta
\rightarrow 0$, $\theta_0 \rightarrow \pi/2$. Both $E$ and $J$ diverge
as $\ln \eta$ at leading order while their difference stays finite:
 \begin{equation} E - J\rightarrow {2 R^2\over \pi \alpha'}= 
 2{\sqrt \lambda\over \pi}
\ .
\end{equation}
 This is to be contrasted with the string spinning in $AdS_5$: there,
$E-S \sim \sqrt\lambda \ln S$, while here $E-J \sim \sqrt\lambda$.
We believe that the logarithm is due to the many gauge-covariant
derivatives that carry the spin $S$ in the dual gauge theory.

As for the strings spinning in $AdS_5$, it seems likely that the
strings spinning in $S^5$ can decay into BPS or almost BPS states:
physically, this corresponds to the string breaking and descending
from the north pole to the equator.  Such processes can be suppressed
by taking $g_{\rm str} \ll 1$.  Also in analogy to strings spinning in
$AdS_5$, we may continue past the limit $\omega < 1$, but the
solutions again change character: instead of a string doubled back on
itself, and stretched out because of its angular momentum, we must now
consider a string stretched around a great circle of $S^5$, and
rotating so that its velocity at the equator of $S^5$ is $\omega$.
Such states are unstable toward ``slipping off the side'' of the
$S^5$.  The relation \thetap\ still holds, and the energy and angular
momentum of these great-circle states may be expressed as
  \eqn{EJGreat}{
   E &= 4 {R^2\over 2 \pi \alpha'} \int_0^{2\pi} d\theta \, 
    \left( 1 - \omega^2 \sin^2\theta \right)^{-1/2} \,,\cr
   J &= 4 {R^2\over 2 \pi \alpha'} \omega \int_0^{2\pi} d\theta \, 
    \left( 1 - \omega^2 \sin^2\theta \right)^{-1/2} \sin^2\theta \,.
  }
 As usual, the integrals can be done in terms of elliptic or
hypergeometric functions.

\subsection{Oscillating String}
\label{Oscillate}

Now let us consider a closed string which oscillates around the center
of $AdS_5$. It first expands from the center of $AdS_5$, reaches a
maximum $\rho$ and then recontracts back to zero; then this motion
repeats.  For such a string, $\phi=\sigma$ and $\rho=\rho (\tau)$.

The space-time energy is a constant of motion:
\begin{equation} E = {R^2\over \alpha'} {dt\over d\tau}\cosh^2 \rho
\ .
\end{equation}
 {}From $T_{++}= T_{--} = 0$ we find
 \begin{equation} (\dot \rho)^2 + \sinh^2 \rho - e^2 \cosh^{-2} \rho = 0
\ ,
\end{equation}
where $e= E\alpha'/R^2$.

The maximum radius of the string is found from 
\begin{equation} \sinh^2 (2 \rho_0) = 4 e^2
\ .
\end{equation}

Presumably these strings correspond to highly excited sigma model
operators.
It is fairly clear that the excitation level is related to the
amplitude $\rho_0$. When $E/\sqrt\lambda $ is small 
we find harmonic oscillations of amplitude $\rho_0$.
Then $\rho_0^2$ should be proportional to the excitation level of the
oscillation:
\begin{equation} \rho_0^2 \sim {n\over \sqrt \lambda} 
\ ,
\end{equation}
 which insures the validity of the usual AdS/CFT relation between
$\Delta$ and the level.  It is less clear how to extrapolate to large
$E/\sqrt\lambda $ where the oscillations are no longer small, and it
is also mysterious to us precisely what gauge theory operators these
states correspond to.

\subsection{Arbitrary conformal dimension}
\label{ArbitraryH}

As a formal exercise, we can generalize the results of
section~\ref{Regge} to solutions that don't obey the constraints, but
rather have $T_{++} = T_{--} = \delta-1$ for some arbitrary conformal
dimension $\delta$.  In this way we hope to get an asymptotic formula
of the type $\alpha \delta = f(\alpha\Delta,\alpha S)$, for the
spinning string in $AdS_5$.  Such a formula is an approximate scaling
form, good when $\alpha \ll 1$ but $\alpha\delta$, $\alpha\Delta$, and
$\alpha S$ are finite.  Equation~\ref{logES} is recovered by setting
$\delta=1$.

{}From $T_{++}=T_{--}=\delta-1$, we arrive at a slight modification of
our previous treatment of the string spinning in $AdS_5$: now
  \begin{equation}
   \left( d\rho \over d\sigma \right)^2 = {4(\delta-1) \over R^2} + 
    \cosh^2\rho - \omega^2 \sinh^2\rho \,,
  \end{equation}
 but the expressions \EnergyC\ and \SpinC\ for the energy and spin are
unchanged.  The integrals can still be done in terms of hypergeometric
functions.  The result is best expressed be defining
  \begin{equation}
   s = {2\pi\alpha' \over R^2} S \qquad 
   e = {2\pi\alpha' \over R^2} E \qquad
   a = \sqrt{1 + 4(\delta-1)/R^2} \,.
  \end{equation}
 Then we obtain
  \begin{equation}\eqalign{
   s &= {\pi\omega \over \sqrt{\omega^2-1}} {a^2 \over \omega^2-1}
     {}_2F_1 \left( {1 \over 2}, {3 \over 2}; 2; 
       {a^2 \over \omega^2-1} \right) \approx 
       {a \over \eta} + {1 \over a} \ln\eta  \cr
   e &= {2\pi \over \sqrt{\omega^2-1}}
     {}_2F_1 \left( -{1 \over 2}, {1 \over 2}; 1; 
       {a^2 \over \omega^2-1} \right) \approx
       {a \over \eta} - {1 \over a} \ln\eta \,,
  }\end{equation}
 where as usual, $\omega = 1+2\eta$.  Solving for $a$, one finds
  \eqn{aSolve}{
   a \approx {2 \over e-s} \ln(e+s) \,.
  }
 Trivial algebra leads to an explicit though slightly messy expression
for the function $f$ in the asymptotic relation $\alpha\delta =
f(\alpha\Delta,\alpha S)$.  Setting $\delta=1$ is equivalent to $a=1$,
and with this constraint, \aSolve\ leads back to \StateDeltaS.

\section{Logarithms in the perturbative regime}
\label{Perturbative}

Logarithmic scaling violations are a characteristic feature of
perturbative gauge theory in four dimensions, observed, for instance,
for gauge theory operators of large dimension but low
``twist''---where twist is dimension minus spin. 
Their anomalous dimensions are responsible for violations
of the Bjorken scaling in deep inelastic scattering
\cite{GrossWilczek}. As an example,
consider the operator
  \begin{equation}
   {\cal O}^\phi_{\mu_1 \ldots \mu_n} = \phi^* 
    \nabla_{(\mu_1} \cdots \nabla_{\mu_n)|} \phi \,,
  \end{equation}
 for a gauge theory with some scalar fields transforming in a given
representation (which we shall eventually specialize to the adjoint,
as appropriate to ${\cal N}=4$ supersymmetric Yang-Mills theory).  The
symmetrization $(\ldots)|$ denotes also a removal of traces.  We
restrict to even $n$ since this is the only non-zero case for ${\cal
N}=4$ SYM.  The gauge-covariant derivative is $\nabla_\mu =
\partial_\mu + i g_{YM} A_\mu$.  In the following calculation, we will
not attempt to construct conformal primaries---so in principle there
can be mixing between the operators we consider and others of twist
two.  Nor will we try to write down the one-loop amplitudes with all
factors and signs precise: the main point will be to remind the reader
of why the dominant one-loop effects grow logarithmically with $n$.
In the free theory, the operator ${\cal O}^\phi$ has dimension $n+2$
and spin $n$, hence twist $2$.  We would like to inquire what
anomalous dimension arises at one loop.  This is closely analogous to
a classic perturbative QCD calculation \cite{GrossWilczek}.  The
dominant graph has two scalars and a gauge boson coming out of the
vertex, and the gauge boson rejoining one scalar line.  It is
convenient to introduce
  \begin{equation}
   {\cal O}^\phi_\Delta = {\cal O}^\phi_{\mu_1 \ldots \mu_n} 
   \Delta^{\mu_1} \cdots \Delta^{\mu_n}
  \end{equation}
 for any vector $\Delta^\mu$.  The anomalous dimension can be read off
from the logarithmic term in the amplitude for the dominant graph:
  \begin{equation}
   {\cal M} = i g_{YM}^2 C_2(R) \int {d^4 p \over (2\pi)^4} 
    {1 \over p^2 (p-k)^2} \sum_{j=0}^{n-1} (\Delta \cdot p)^j
    (\Delta \cdot k)^{n-j} = c_\phi (\Delta \cdot k)^n 
    \ln\Lambda^2 + \ldots \,,
  \end{equation}
 where $C_2(R)$ is the Casimir of the representation in which $\phi$
falls.  The $j^{\rm th}$ term in the sum comes from the term
  \begin{equation}
   ig_{YM} \phi^* (\partial_{\mu_1} \cdots 
    \partial_{\mu_j} A_{\mu_{j+1}}) \partial_{\mu_{j+2}} \cdots
    \partial_{\mu_n} \phi
  \end{equation}
 in ${\cal O}^\phi_{\mu_1 \ldots \mu_n}$.  We still intend the
traceless symmetrization of $\mu_1 \ldots \mu_n$, though it is no
longer explicit in the notation.  To kill off all contact terms
and pick up only the logarithmic term that rescales ${\cal
O}^\phi_{\mu_1 \ldots \mu_n}$, we differentiate $n$ times with respect
to $k$ and then set $k=0$:
  \begin{equation}\eqalign{
   &i g_{YM}^2 C_2(R) \int {d^4 p \over (2\pi)^4} \sum_{j=0}^{n-1}
    \left( {n \atop j} \right) {2^j j! \over p^{2j+4}}
    p_{\mu_1} \cdots p_{\mu_j} (\Delta \cdot p)^j (n-j)!
    \Delta_{\mu_{j+1}} \cdots \Delta_{\mu_n}  \cr
   &\qquad{} = c_\phi n! \Delta_{\mu_1} \cdots \Delta_{\mu_n} 
     \ln\Lambda^2 \,.
  }\end{equation}
 The integral diverges logarithmically.  Replacing $\int_0^\infty {dp
\over p}$ by $\ln\Lambda$, we wind up with
  \begin{equation}
   {i g_{YM}^2 C_2(R) \over 16\pi^2} 
    {1 \over 2\pi^2} \int_{S^3} d^3 \Omega \sum_{j=0}^{n-1} 
    2^j \hat{p}_{\mu_1} \cdots \hat{p}_{\mu_j} 
     (\Delta \cdot \hat{p})^j \Delta_{\mu_{j+1}} \cdots \Delta_{\mu_n}
     = c_\phi \Delta_{\mu_1} \cdots \Delta_{\mu_n} \,,
  \end{equation}
 where $\hat{p}_\mu$ is a unit vector in the unit three sphere, $S^3$,
in Euclidean momentum space.  The removal of trace terms can be
accomplished, heuristically, by taking $\Delta_\mu$ to be null.  The
angular averaging over $S^3$ that is left is a type of integral
explained in \cite{GrossWilczek}.  The result of doing this integral is
  \begin{equation}
   {i g_{YM}^2 C_2(R) \over 16\pi^2} \sum_{j=0}^{n-1} 
    {2^j j! \over (2j+2)!!} \Delta_{\mu_1} \cdots \Delta_{\mu_n} = 
    c_\phi \Delta_{\mu_1} \cdots \Delta_{\mu_n} \,,
  \end{equation}
 where $n!! = \prod_{j=0}^{\lfloor n/2 \rfloor} (n-2j)$.  Simplifying,
we obtain
  \begin{equation}
   c_\phi = {i g_{YM}^2 C_2(R) \over 32\pi^2} \sum_{j=1}^n {1 \over j} 
    \sim {i g_{YM}^2 C_2(R) \over 32\pi^2} \ln n
  \end{equation}
 for large $n$.  This shows that the wavefunction renormalization, and
hence the anomalous dimension, do scale as $\lambda \ln n$ for large
$n$, as desired.\footnote{We thank T.~Petkou for pointing out to us
that a similar result has been obtained in \cite{do}.}
  Calculations of this sort serve as classic
demonstrations of logarithmic scaling violation in deep inelastic
scattering.  It is fascinating to see logarithms also in the
supergravity dual, which is far from the perturbative regime!  Of
course, here the scaling is $\sqrt{\lambda} \ln n$.  The turnover
from $O(\lambda)$ to $O(\sqrt\lambda)$ behavior is reminiscent of the
coefficient for Wilson loops \cite{juanWilson,esz,GrossDrukker}.
For consistency with
our supergravity result, the $k$-loop correction
to the anomalous dimension should grow for large $S$ as 
$\lambda^k \ln S$. This behavior is indeed observed in both supersymmetric
and non-supersymmetric gauge theories \cite{Korch}.
The successes \cite{esz,GrossDrukker} in computing
circular Wilson loops to all orders in the 't~Hooft coupling lead us
to hope that similar exact resummations (perhaps of rainbow graphs)
might lead to a direct comparison between strongly
coupled field theory and string theory near the leading edge of the
Regge trajectory.  This line of thought is particularly attractive in
light of the fact that the states we are considering are nearly
back-to-back Wilson loops---compare figure~\ref{figA}b)
to~\ref{figA}c).

We should emphasize that the conclusion that 
in ${\cal N}=4$ SYM the $k$-loop correction to the anomalous
dimension grows at most as $\ln S$ is highly non-trivial. 
In gauge theories individual graphs at $k$-loop order 
grow as fast as $\lambda^k (\ln S)^{2k-1}$ \cite{Gross}.
Often there are cancellations, however.\footnote{We thank 
E.~Witten for pointing this out to us.}
Analyses of
the two-loop \cite{Floratos,Lop}, and all higher order corrections \cite{Korch}
indeed indicate that the anomalous dimensions grow only as $\ln S$
perturbatively, and probably also non-perturbatively. 
Remarkably, there seems to be little difference in the
behavior of these anomalous dimensions between the 
${\cal N}=4$ SYM theory and QCD. 
This should have an interesting effect
on the behavior of the deep inelastic scattering.

We should also note that there exist simple bilocal operators
whose expansion generates all higher spin operators introduced
above:
  \eqn{biloc}{
   &\Tr\left\{ \phi^* (x)\ {\rm Pexp} (i\int_x^y A_\mu dx^\mu)\ \phi(y)
    \ {\rm Pexp} (-i\int_x^y A_\mu dx^\mu) \right\}  \cr &\qquad\qquad =
   \sum_{n=0}^\infty {\cal O}^\phi_{\mu_1 \ldots \mu_n} 
    (x-y)^{\mu_1} \cdots (x-y)^{\mu_n} \,.
  }
Thus,
if the separation between $x$ and $y$ is taken to be light-like
then expanding in powers of $(x-y)^\mu$ we recover precisely the
higher spin operators made out of two scalar fields and some number
of covariant derivatives that we have been studying. 
Similar bilocal operators can be made by
replacing $\phi$ by a fermion field or $F_{\mu\nu}$.
The growth of the anomalous dimensions with $S$ is clearly related to
the presence of the path-ordered exponentials in (\ref{biloc}),
\cite{Korch}.
It is also tempting to think 
that the ``folded string'' nature of the bilocal operators
is related to the spinning string picture of the dual states in
AdS$_5$.

\section{Discussion}

Our discussion of spinning string solitons has so far been phrased
in terms of the global AdS$_5$ metric of Lorentzian signature.
If we continue to Euclidean signature, both in spacetime and on the
world sheet, then what we have done can be rephrased in terms of
the Euclidean $AdS_5$ space written in Poincar\' e coordinates.

Introducing coordinate $r =\sinh \rho$ in (\ref{global}), we may 
write the 
Euclidean continuation of the metric as
\be \label{newglob}
ds^2 = (1+r^2) dt^2 + {dr^2\over 1+r^2} + r^2 d\Omega_3^2
\ .
\ee
This may be brought to the form
\be
ds^2 = d\varphi^2 + e^{2\varphi}\sum_{i=1}^4 dx_i^2 =
d\varphi^2 + e^{2\varphi} (dy^2 + y^2 d\Omega_3^2)
\ee
by the coordinate change
\be \varphi = t +{1\over 2}\ln (1+r^2)\ ,
\qquad y= {r\over\sqrt{1+r^2}} e^{-t}\ .
\ee
Now we may substitute the Euclidean continuation of the
spinning string soliton studied in section 3 to derive a soliton
in Poincar\' e coordinates.

It is of further interest to consider more general metrics of the form
\be
ds^2 = d\varphi^2 + a^2(\varphi)\sum_{i=1}^4 dx_i^2
\ .
\ee
which are relevant to more general gauge theories.
In this case finding a spinning string soliton solution
in closed form is less straightforward. If we attempt to
transform to $S^3$ slicing of the geometry, then we generally
arrive at a time-dependent metric of the form
\be f(r,t) dt^2 + g(r,t) dr^2 + r^2 d\Omega_3^2
\ .
\ee
The time-dependence in these coordinates is a sign of broken conformal 
invariance. Suppose, however, that in the UV (for large $\varphi$)
the metric is asymptotic to AdS$_5$: $a(\varphi)\rightarrow 
e^{\varphi}$.
Then in the $S^3$ slicing of the geometry the metric approaches
(\ref{newglob}) for large $r$. For a spinning string with very large
spin $S$, the dominant contributions to conformal dimension and
spin arise from this large $r$ region where the metric is approximately
AdS$_5$. We conclude, therefore, that the relation
(\ref{StateDeltaS}) applies in the limit of large $S$ for all gauge theories
that approach the ${\cal N}=4$ SYM theory asymptotically in the
UV. We hope to return to construction and study of solitons
in theories with different UV asymptotics in a future publication.

In this paper we have shown that some characteristically stringy effects
can be extracted from the AdS/CFT duality by studying states whose
Lorentz spin $S$ is much greater than $R^2/\alpha'=\sqrt\lambda$.
The dual description of gauge theory operators with such large $S$ cannot
be given in terms of local fields in AdS$_5$; instead it involves
strings stretched across AdS$_5$ in such a way that they probe
the metric near its boundary. Our results are reliable for
$R^2/\alpha' \gg 1$, but one could hope that they are applicable qualitatively
even for $R^2/\alpha'$ of order 1. 

In fact, recently 
there appeared interesting ideas on how to extend the AdS/CFT
duality to large $N$ gauge theory at weak `t Hooft coupling
$\lambda$ \cite{Vas,Sundborg,polWord,Ed,Sezgin,Mikh}.
In this case the relation between $R^2/\alpha'$ and $\lambda$ is not known
precisely. Naively, $R^2/\alpha'$ is of order $\lambda$ in the weak
coupling limit, but it is also possible that it approaches a
number of order $1$ (there might not be an unambiguous
definition of $R$ when it becomes as small as $\sqrt{\alpha'}$). 
If we suppose that our results may be
extrapolated to this situation, then the dual description of
states with $S\gg 1$ is not in terms of local fields in $AdS_5$
but rather in terms of non-local stretched string states.
This comment could be relevant to the search for infinite number of
higher spin symmetries at $\lambda=0$
because the gauge theory operators we are discussing 
are precisely the higher spin currents.

Our results suggest that extra conservation laws, obvious at
$\lambda=0$, persist at non-zero coupling in some asymptotic sense.
Since we have $(\Delta-S)/S \ll 1$ for operators on the leading
Regge trajectory, we find that the spacing of dimensions of
these operators is asymptotically integral.  These facts are not
enough to predict an exact conservation law, as one may be able to do
if $\Delta=S+2$.  But they are suggestive of asymptotic conservation
laws as $S$ tends to $\infty$.

\section*{Acknowledgments}
We are very grateful to J. Maldacena for collaboration at early stages
of this project.
We also thank B. Altshuler, M. Bertolini, K. Efetov, G. Korchemsky, 
P. Ouyang and E. Witten for useful discussions.
The work of SSG is supported in part by 
DOE grant DE-FG02-91ER40671 and an Outstanding Junior Investigator award.
The work of IRK and AMP is supported in part by the NSF Grant
PHY-9802484.

\begingroup\raggedright\endgroup


\end{document}